\documentclass{article}
\usepackage{spconf,amsmath,graphicx,booktabs,amsfonts}
\usepackage{caption}
\usepackage{subcaption}
\usepackage{tabularx}

\newcolumntype{b}{X}
\newcolumntype{s}{>{\hsize=.75\hsize}X}

\title{Discriminative Speech Recognition Rescoring with Pre-trained Language Models}
\name{Prashanth Gurunath Shivakumar, Jari Kolehmainen, Yile Gu, Ankur Gandhe, Ariya Rastrow, Ivan Bulyko}
\address{Amazon Alexa AI, USA\\
\small{\texttt{\{psshvak,jkolehm,yilegu,aggandhe,arastrow,ibbulyko\}@amazon.com}}}
\copyrightnotice{979-8-3503-0689-7/23/\$31.00~\copyright2023 IEEE}
\begin{document}
%\ninept
%
\maketitle
\begin{abstract}
Second pass rescoring is a critical component of competitive automatic speech recognition (ASR) systems. Large language models have demonstrated their ability in using pre-trained information for better rescoring of ASR hypothesis. Discriminative training, directly optimizing the minimum word-error-rate (MWER) criterion typically improves rescoring. In this study, we propose and explore several discriminative fine-tuning schemes for pre-trained LMs. We propose two architectures based on different pooling strategies of output embeddings and compare with probability based MWER. We conduct detailed comparisons between pre-trained causal and bidirectional LMs in discriminative settings. Experiments on LibriSpeech demonstrate that all MWER training schemes are beneficial, giving additional gains upto 8.5\% WER. Proposed pooling variants achieve lower latency while retaining most improvements. Finally, our study concludes that bidirectionality is better utilized with discriminative training.

%Second pass rescoring is a critical component of most state-of-the-art automatic speech recognition (ASR) systems.
%Large language models (LLM) have demonstrated their ability in using pre-trained information for better rescoring of ASR hypothesis.
%Further, discriminative training, directly optimizing the minimum word-error-rate (MWER) criterion typically improves rescoring.
%In this study, we propose and explore several discriminative fine-tuning schemes for pre-trained language models.
%We propose two architectures based on different pooling strategies of output embeddings and compare them with typical probability based MWER.
%We also conduct detailed comparisons between pre-trained causal (GPT-2) and bidirectional (BERT) language models with comparable sizes in discriminative settings.
%We conduct experiments on LibriSpeech corpus using publicly available ASR and LMs, and demonstrate that all MWER training schemes are beneficial, with additional 5.9\% WER relative improvement for GPT-2 model and 8.5\% WER relative for BERT.
%Compared with typical probability-based models, the proposed pooling variants achieve lower latency while retaining most of the accuracy improvements, especially in case of BERT.
%Finally, our study concludes that bi-directionality is better utilized with discriminative training i.e., difference between GPT-2 and BERT is larger after MWER training.
\end{abstract}
\section{Introduction}
\label{sec:intro}
End-to-end speech recognition systems have seen tremendous advancements in making speech recognition more accurate and closing the gap to human levels.
However, large models with all neural end-to-end training requires large amount of transcribed speech data to achieve better performance.
Availability of labelled speech data is always limited compared to the largely, freely available text data.
%The ability of language models to ingest this data to effectively learn human language makes second pass rescoring particularly effective in these scenarios.
Meanwhile, the recent advancements in computational efficiency has enabled scaling language models.
The large language models (LLM) have demonstrated their ability to ingest vast amount of text data to match almost human level performance in many benchmark tasks.
Exploiting knowledge from these LMs to improve second pass rescoring has become particularly attractive.

Several prior works have explored employing pre-trained models such as GPT, BERT for rescoring \cite{huang2019empirical,shin2019effective, udagawa2022effect,salazar2020masked, xu2022rescorebert,chen2023large, shivakumar2023distillation,kolehmainen2023personalization,gu2023scaling}.
\cite{huang2019empirical} conducted extensive study for rescoring with GPT models.
\cite{shin2019effective} proposed using BERT for utterance level ASR rescoring and demonstrated the importance of bi-directional encoding. 
Pseudo-log-likelihood (PLL) for each n-best hypothesis is computed by summing the log-likelihoods obtained from the model by masking each token position in the input sequence, one-by-one. 
The PLL is interpolated with the first pass score to obtain the final n-best rank.
Given that the PLL computation is expensive, \cite{salazar2020masked} further extended BERT for mask-less scoring where the [CLS] token is fine-tuned using L2-loss to predict the PLL.
This dramatically reduces the computation and makes BERT-based rescoring practically feasible.
\cite{salazar2020masked} conducted a detailed comparison of GPT-2 and BERT models for rescoring which demonstrates the effectiveness and advantages of bi-directional encoding of BERT models for rescoring.
\cite{udagawa2022effect} conducted empirical study of GPT-2 and BERT models for rescoring on state-of-the-art first pass model and evaluates the effect of additional context.
LLMs ranging from 70M to 540B parameters for second pass rescoring is investigated in \cite{chen2023large} with findings of significant wins over strong first pass model.

On the other hand, several works have demonstrated the effectiveness of discriminative training with MWER criterion \cite{prabhavalkar2018minimum, chiu2018state, 7472827, gandhe2020audio,hu2021transformer,shivakumar2023distillation,kolehmainen2023personalization,gu2023scaling}.
\cite{prabhavalkar2018minimum,chiu2018state} proposed MWER loss for training end-to-end ASR with encoder-decoder architectures.
\cite{7472827} trained RNN-LMs with MWER for rescoring and achieved improvements over using log-likelihoods.
\cite{gandhe2020audio} used MWER training with LSTM language models attending to audio embeddings from first-pass encoder and showed improvement over cross-entropy based training.
In \cite{hu2021transformer}, a transformer based deliberation rescorer attending to first pass audio encoding and ASR hypothesis encoding is trained with MWER criteria and reported gains of 9\% in WER.

Despite the advantages of MWER training, there are very few studies on adding MWER training to further improve rescoring performance of a pre-trained model.
%pretrained models are in limelight for ASR rescoring mainly due to large pre-training and the benefits it brings.
%More recently, 
\cite{xu2022rescorebert} proposed MWER training for BERT models, which is the first work that combines the importance of large pre-training with discriminative learning.
The work builds upon the findings from \cite{shin2019effective} and \cite{salazar2020masked} by proposing a technique that uses a pre-trained BERT model with a feed-forward layer on the [CLS] embedding similar to \cite{salazar2020masked} and further extends by fine-tuning with MWER loss.
Significant gains are reported with discriminative training over the n-best hypotheses \cite{xu2022rescorebert,shivakumar2023distillation,kolehmainen2023personalization,gu2023scaling}.
The scheme proposed in \cite{xu2022rescorebert} is an approximation and deviates from typical sequence probability based computation in favor of latency.

%While the advantages in combining the benefits of large pre-training and discriminative training is evident, there has been no prior work and a good sense of understanding on the interplay between MWER training and bi-directionality with large pre-trained models.
However, there lacks a work on combining MWER training with a pre-trained causal LM, and a comparison with its bidirectional counterpart.
Although bi-directional encoding could be important, causal language models have certain advantages with latency and also naturally fits into streaming frameworks that make it attractive.
For example, with causal models, re-scoring need not always wait until the completion of first pass.
Moreover, promising performance of causal language models with zero-short learning \cite{brown2020language} and scaling \cite{kaplan2020scaling} attracts interest in its applicability to rescoring \cite{li2023prompting}.

In this study, we attempt to bridge this gap in understanding how to effectively combine MWER training with both causal and bidirectional pre-trained models.
%the role of pre-training, bidirectional encoding and MWER training in application to ASR rescoring.
The contribution of this paper are in two folds.
First, we propose three different ways to fine-tune LLM with MWER criterion. 
We explore three techniques: (i) using last-token embedding from GPT-2, (ii) attention pooling on all tokens, and (iii) sequence probability based MWER criteria \cite{gandhe2020audio}.
Second, we perform a detailed comparison between pre-trained causal language model, i.e., GPT-2, and bidirectional model, i.e., BERT to assess the role of bidirectional encoding in ASR rescoring in MWER discriminative training setting.
To the best of our knowledge this is the first work to explore pretrained, decoder-only transformer based causal language models with MWER training for ASR rescoring.

The rest of the paper is organized as follows: Section~\ref{sec:sec2} describes the proposed techniques for discriminative training GPT-2 models with MWER loss. Section~\ref{sec:data_exp_setup} presents the datasets and the experimental setup. Results are discussed in Section~\ref{sec:results} and finally conclusions are drawn in Section~\ref{sec:conclusion}.

%\section{ASR Rescoring Models}
\label{sec:sec2}

\begin{figure*}[t]
    \centering
    \begin{subfigure}[t]{0.4\textwidth}
        \centering
        \includegraphics[width=0.75\linewidth]{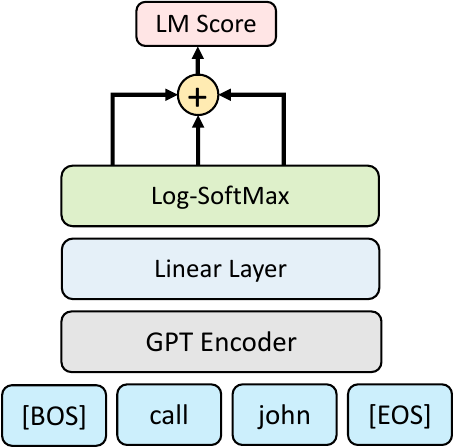}
        \caption{GPT-2}\label{fig:gpt}
    \end{subfigure}
    \begin{subfigure}[t]{0.55\textwidth}
        \centering
        \includegraphics[width=\linewidth]{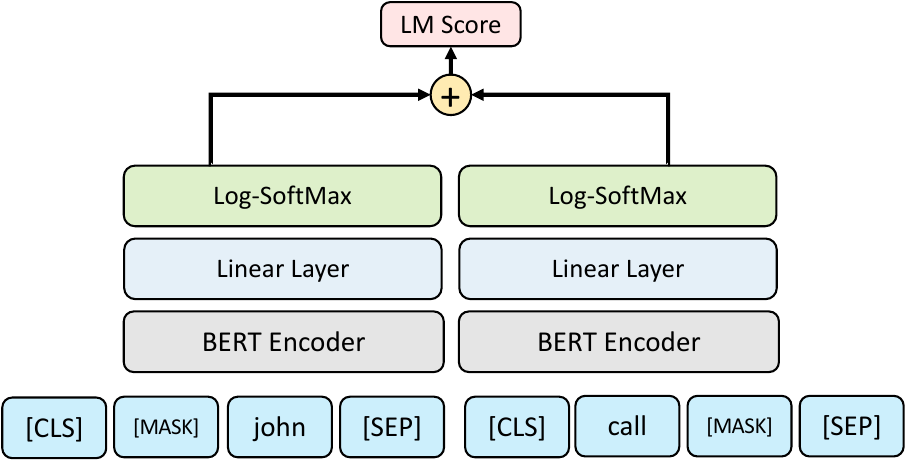}
        \caption{BERT}\label{fig:bert}
    \end{subfigure}
    \medskip
    \vspace{5mm}
    \begin{subfigure}[t]{0.4\textwidth}
        \centering
        \includegraphics[width=0.75\linewidth]{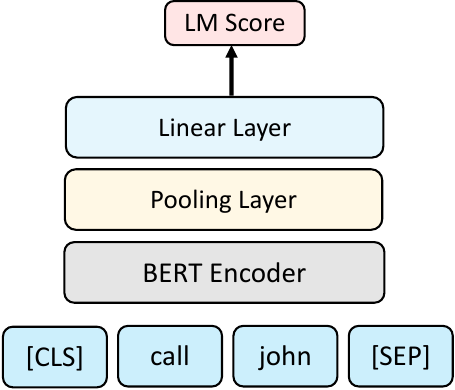}
        \caption{RescoreBERT}\label{fig:rescorebert}
    \end{subfigure}
    \begin{subfigure}[t]{0.4\textwidth}
        \centering
        \includegraphics[width=0.75\linewidth]{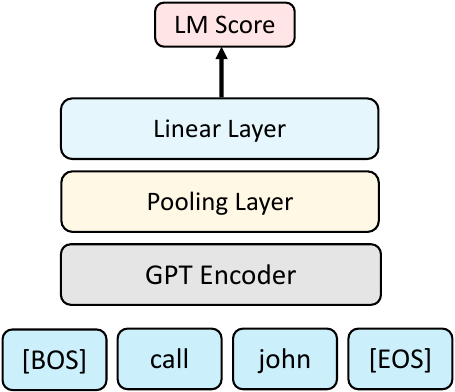}
        \caption{RescoreGPT}\label{fig:rescoregpt}
    \end{subfigure}
    \caption{Illustration of model architectures (a): GPT-2 based rescoring where token probabilities are predicted for tokens after beginning of sentence by Eq.~\eqref{eq:gpt_loss}, (b): BERT rescoring where each token is predicted by the masked sequence according to Eq.~\eqref{eq:pll}, (c) RescoreBERT that uses pooling layer on the BERT encoder and predicts a score using a linear layer, (d): RescoreGPT model that uses a pooling layer on the GPT-2 encoder and a linear layer to predict score. Note, for architectures (c) and (d), the pooling layer is either CLS embedding or attention pooling over all token embeddings.}
    \label{fig:fig}
\end{figure*}

\section{Non-Discriminative ASR Rescoring}\label{sec:sec2}
Likelihood scores from pre-trained language models can be natively used to rescore ASR n-best by interpolating with the first pass acoustic model scores:
\begin{equation}\label{eq:interpolate}
    s_i = log P_{LM}(y_i) + \lambda log P_{AM}(x|y_i)
\end{equation}
where $P_{AM}(x|y_i)$ is the sequence probability of the 1st pass acoustic model, given an input audio sequence, $x$, for $i^{th}$ ASR hypothesis, $y_i$, $P_{LM}(y_i)$ is the likelihood of the rescoring language model and $\lambda$ is the interpolation weight.

\subsection{GPT-2}
In case of GPT-2, the likelihood for a sequence of length $L$ is computed in a auto-regressive fashion by modeling:
\begin{equation}\label{eq:gpt_loss}
    P_{LM}(y_i) = \prod_{t=1}^{L}P(y_{i,t}|y_{i,1},\ldots,y_{i,t-1})
\end{equation}
using a soft-max over the entire vocabulary.

\subsection{BERT}
In contrast, bi-directional encoding models such as BERT use masked language modeling where log likelihood for sentences are not directly available.
However, several works have demonstrated pseudo-log-likelihood (PLL) scores are a viable replacement \cite{shin2019effective} for rescoring.
PLL is computed as 
\begin{equation}\label{eq:pll}
    PLL(y) = -\sum_{t=1}^L \mathrm{log} P(y_t|y_{\backslash t})
\end{equation}
where $y_{\backslash t} = \{y_1\ldots, y_{t-1},[MASK],y_{t+1},\ldots, y_L\}$.
However, this comes with the expense of increased computation and memory since it requires copies and inference passes equal to the length of the sequence (see Fig.\ref{fig:bert}).

\section{Discriminative ASR Rescoring}
Authors in \cite{prabhavalkar2018minimum} propose to directly minimize expected word error rate while training sequence-to-sequence attention based models for ASR.
Given an input audio sequence $x$ with ground-truth transcript $y^*$ and ASR hypothesis $y$, MWER loss can be computed as:
\begin{equation}
    L_{mwer}(x, y^{*}) = \mathbb{E}[\mathcal{E}(y,y^{*})] = \sum_{y}P(y|x)\mathcal{E}(y,y^{*})
\end{equation}
where $\mathcal{E}(y, y^*)$ is the edit distance between the ASR hypothesis, $y$, and groundtruth transcript, $y^*$, $P(y|x)$ is the probability of ASR hypothesis $y$ given the input audio sequence $x$. 
To make the loss tractable, the expectation is approximated by restricting the sequence probability over the n-best hypothesis.
\begin{equation}\label{eq:mwer}
    L_{mwer}(x, y^*) =  \sum_{i=1}^{N}P(y_i|x)\mathcal{E}(y_i,y^*)
\end{equation}
where $N$ refers to N-best ASR hypotheses.

\subsection{Sequence probability based MWER}\label{sec:ll}

The above MWER loss for discriminatively training can be applied for 2nd pass rescoring model \cite{gandhe2020audio}.
For rescoring, the n-best posterior probability $P(y_i|x)$ is computed by considering a linear combination of first pass acoustic model score and the LM, $s_i$:
\begin{equation}
    P(y_i|x) = \frac{e^{s_i}}{\sum_{j=1}^{N}e^{s_j}}
\end{equation}

\subsubsection{GPT + MWER}
Similarly, we can use GPT-2 to compute MWER loss as follows:
\begin{equation}\label{eq:gpt_mwer}
    L_{mwer}(x, y^*) =  \sum_{i=1}^{N}\frac{e^{s_i}}{\sum_{j=1}^{N}e^{s_j}}\mathcal{E}(y_i,y^*)
\end{equation}
and
\begin{equation}
    s_i = log (\prod_{t=1}^{L}P(y_{i,t}|y_{i,1},\ldots,y_{i,t-1})) + \lambda log P_{AM}(x|y_i)
\end{equation}
where $s_i$ and $s_j$ are computed by plugging Eq.~\eqref{eq:gpt_loss} into Eq.~\eqref{eq:interpolate}.
The model architecture is depicted in Fig.~\ref{fig:gpt}.

\subsubsection{BERT + MWER}
In case of BERT, we propose to use Eq.~\eqref{eq:pll} as LM-score, i.e., $s_i$ is given by:
\begin{equation}
    s_i = -\sum_{t=1}^L \mathrm{log} P(y_t|y_{\backslash t}) + \lambda \mathrm{log} P_{AM}(x|y_i)
\end{equation}
which is plugged into Eq.~\eqref{eq:gpt_mwer} for MWER loss computation.
The model architecture is depicted in Fig.~\ref{fig:bert}.

\subsection{Embedding based MWER Rescoring}
As demonstrated in \cite{salazar2020masked} and \cite{xu2022rescorebert} for BERT, an alternative way to compute a score for rescoring is to extract [CLS] embedding for the sentence and add a linear layer.
We name these model architectures RescoreBERT and RescoreGPT.
\subsubsection{RescoreBERT}\label{sec:pll}
\cite{salazar2020masked} showed that the PLLs can be approximated by fine-tuning the classification ([CLS]) token via L2 regression. 
Building upon this, \cite{xu2022rescorebert} proposed MWER training over the [CLS] token, i.e.,
\begin{equation}
    \mathrm{log} P_{LM}(y_i) = W_F(BERT_{CLS}(y_i))
\end{equation}
where $BERT_{CLS}$ refers to the BERT [CLS] embedding and $W_F$ is a learnable feed-forward layer.
The model architecture is illustrated in Fig.~\ref{fig:rescorebert}.
%\begin{equation}
%    L_{mwer}(x, y^*) = \sum_{i=1}^{N}P(y_i|x)\mathcal{E}(y_i,y^*)
%\end{equation}
\subsubsection{RescoreGPT}
In case of GPT-2 models, we propose to use the probability computation on the last token for MWER training.
The motivation here is that by considering the last token embedding, the model has seen all the tokens and has similar information as BERT model.
Thus, we propose to use the following for MWER computation:
\begin{equation}
    \mathrm{log} P_{LM}(y_i) = W_F(GPT_{last}(y_i)) %P(y_{i,L}|y_{i,1},\ldots,y_{i,L-1})
\end{equation}
%where $L$ is the length of the sequence $y_i$.
Note, this has advantages in reducing the memory and computational footprint during training and inference, drastically, since soft-max is computed over n-best and not over the vocabulary, and only once per utterance.
The architecture is illustrated in Fig.~\ref{fig:rescoregpt}, where the pooling layer is the last token embedding.

\subsubsection{Attention Pooling}\label{sec:attn_pooling}
A natural extension of the above approach is to make use of all the output embeddings from GPT instead of only the embedding corresponding to the last token.
One way to achieve this is by attention pooling over the GPT output embeddings, i.e.,
\begin{equation}
    \mathrm{log} P_{LM}(y_i) = W_F(\mathrm{softmax}(\hat{Q}K^T/\sqrt{d_k})V)
\end{equation}
and 
\begin{align*}
    \hat{Q} = \hat{q}W_Q && K = HW_K && V = HW_V \hfill
\end{align*}
where $H=\{h_1,\ldots,h_L\}$ represent hidden output embeddings, $W_Q$, $W_K$, $W_V$ are learnable weights associated with query, key, values respectively, and $\hat{q}$ is a learnable vector. 
This can be adopted similarly in case of both GPT and BERT encoders as shown in Fig.~\ref{fig:rescoregpt} and ~\ref{fig:rescorebert}.

\section{Datasets and Experimental Setup}\label{sec:data_exp_setup}

\subsection{Data}
\label{sec:data}
Publicly available LibriSpeech corpus is employed for experimentation.
The dataset comprises approximately 1000 hours of read English speech derived from audiobooks from LibriVox \cite{7178964}.
We also make use of Librispeech LM corpus based on Project Gutenbert books for domain adaptation \cite{7178964}.

\begin{table*}[t]
    \centering
    \begin{tabularx}{\textwidth}{bssss}
    
    \toprule
    Model & Loss & Latency $\lbrack \mathrm{ms} \rbrack$ & test-clean & test-other \\
    \midrule
    Whisper \cite{radford2022robust} & - & - & 5.67 & 12.88 \\
    \midrule
    GPT-2 & CE & 49 & 4.77 (15.87\%) & 11.19 (13.12\%) \\
    BERT \cite{shin2019effective} &  CE & 6115 & \textbf{4.71 (16.93\%)} & 11.19 (13.12\%) \\
    \midrule
    GPT-2 & MWER & 49 & 4.52 (20.34\%) & 10.89 (15.45\%) \\
    GPT-2 & MWER + CE & 49 & 4.49 (20.79\%) & 10.90 (15.37\%) \\
    %\midrule
    %BERT &  CE & 6115 & \textbf{4.71 (16.93\%)} & 11.19 (13.12\%) \\
    BERT & MWER & 6115 & 4.36 (23.10\%) & 10.93 (15.14\%) \\
    BERT & MWER + CE & 6115 & \textbf{4.31 (23.99\%)} & 10.81 (16.07\%) \\
    \midrule
    RescoreGPT & MWER & 34 & 4.72 (16.75\%) & 11.03 (14.36\%) \\
    \hfill with Attn. pooling & MWER & 35 & 4.65 (17.99\%) & 11.60 (13.35\%) \\    
    \midrule
    RescoreBERT \cite{xu2022rescorebert} & MWER & 33 & 4.37 (22.93\%) &  10.80 (16.15\%) \\
    \hfill with Attn. pooling & MWER & 33 & 4.38 (22.75\%) & \textbf{10.78 (16.30\%)} \\
    \midrule
    10-best Oracle & - & - & 3.63 (35.98\%) & 9.22 (28.42\%) \\ 
    \bottomrule
    \end{tabularx}
    \caption{WER results for different models. Relative \% WER-reduction is shown in the brackets respect to the 1st pass model. Latency numbers present average time spent in milliseconds for rescoring 10 hypothesis of sequence length 64 for various models on a single NVIDIA T4 GPU.}
    \label{tab:results}
    \vspace{-2mm}
\end{table*}

\subsection{Experimental Setup}
\label{sec:exp_setup}
\subsubsection{First-pass ASR}
For all our experiments, we use Whisper tiny model \cite{radford2022robust}.
The architecture of the Whisper first-pass is based on Transformer sequence-to-sequence models \cite{vaswani2017attention}.
Whisper-tiny comprises 39M parameters and is trained in a supervised manner on 680k hours of audio collected from internet.
We generate top-10 hypothesis for second-pass rescoring purposes.
The hypothesis is post-processed by removing punctuation and converting case according to target LibriSpeech transcripts.
The oracle WER of 10-best on test-clean and test-other is 3.63\% and 9.22\% respectively.

\subsubsection{Second-pass Rescoring}
We use the publicly available GPT-2 model \cite{radford2019language} and bert-base-cased as the BERT \cite{devlin2018bert} model for all the experiments.
Since we plan to compare the role of bidirectionality, special care is taken to ensure that the two models are similarly sized (110M parameters in BERT; 117M parameters in GPT-2), pre-trained on comparable data-sets (BERT on BooksCorpus and Wikipedia; GPT-2 on WebText) and the experimental setting is as similar as possible, following \cite{salazar2020masked}.
%The GPT-2 model is a decoder-only model \cite{liu2018generating} based on Transformer architecture \cite{vaswani2017attention}, which uses causal language modeling scheme \cite{radford2018improving} for unsupervised pre-training.
%The model has 124M parameters and is trained on millions of webpages.
%For experiments on BERT, we use publicly available base-bert-cased.
%The BERT contains 109M parameters which makes it suitable for fair comparison with GPT-2.

We domain adapt both BERT and GPT-2 models on the Librispeech LM corpus for all our experiments.
Next, MWER training is performed as described in Section~\ref{sec:sec2}.
Finally, for rescoring, we tune the optimal interpolation weight $\lambda$ in Eq.~\eqref{eq:interpolate} on the corresponding development sets.

Batch size of 16 is used during domain adaptation and batch size of 4 is used during MWER training.
Learning rate schedule is used during training with an initial learning rate of $1e-5$.
Optimal checkpoint is picked on the corresponding dev-sets and the results on unseen test-sets are reported.

\subsection{Baseline}
In this study, we employ two baselines: (i) 
the first-pass, tiny-Whisper model, (ii) Log-likelihood based rescoring for GPT-2, PLL based rescoring for BERT. \cite{shin2019effective}.
For vanilla log-likelihood and PLL rescoring baselines, the BERT and GPT-2 models are domain adapted first on librispeech LM corpus and then on audio corpus for fair comparison.
%Similarly, in case of RescoreBERT, we first domain adapt base-bert-cased on LibriSpeech LM corpus and finally fine-tuning is performed on the LibriSpeech audio data with MWER objective as described under Section~\ref{sec:pll}.

%Results are reported in terms of absolute word error rate and relative improvements over the first pass baseline.

\section{Results}
\label{sec:results}

%\begin{table*}[t]
%    \centering
%    \begin{tabular}{llll}
%    \toprule
%    Model & Loss & test-clean & test-other \\
%    \midrule
%    Whisper \cite{radford2022robust} & - & 5.67 & 12.88 \\
%    GPT-2 & CE & 4.77 (15.87\%) & 11.19 (13.12\%) \\
%    BERT &  CE & \textbf{4.71 (16.93\%)} & 11.19 (13.12\%) \\
%    \midrule
%    RescoreBERT \cite{xu2022rescorebert} & MWER  & 4.37 (22.93\%) &  10.80 (16.15\%) \\
%    \hfill + Attn. pooling & MWER & 4.38 (22.75\%) & \textbf{10.78 (16.30\%)} \\
%    RescoreGPT & MWER & 4.72 (16.75\%) & 11.03 (14.36\%) \\
%    \hfill + Attn. pooling & MWER & 4.65 (17.99\%) & 11.60 (13.35\%) \\
%    GPT & MWER & 4.52 (20.34\%) & 10.89 (15.45\%) \\
%    GPT & MWER + CE & 4.49 (20.79\%) & 10.90 (15.37\%) \\
%    BERT & MWER & & \\
%    BERT & MWER + CE & \textbf{4.31 (23.99\%)} & 10.81 (16.07\%) \\
%    \midrule
%    10-best Oracle & - & 3.63 (35.98\%) & 9.22 (28.42\%) \\ 
%    \bottomrule
%    \end{tabular}
%    \caption{WER Results (Relative \% WER-reduction).}
%    \label{tab:results}
%\end{table*}

Table~\ref{tab:results} presents the experimental results of the baselines and proposed MWER discriminative training schemes on GPT-2 and BERT models.
The Whisper first pass gives 5.67\% and 12.88\% WER on the test-clean and test-other sets of LibriSpeech.
Typical log-likelihood based rescoring provides upto 15.9\% improvement over the baseline for GPT-2, after domain adaptation on LibriSpeech.
PLL based rescoring \cite{shin2019effective} with BERT outperforms GPT-2 giving upto 16.9\% relative improvement over the first-pass.
Improvements with BERT (1.26\%) relative to GPT-2 can be attributed to the bi-directional encoding and is consistent with \cite{shin2019effective,salazar2020masked}.

%The bottom half of Table~\ref{tab:results} lists results with discriminative training.
%The rows with Attn. pooling refers to models trained using attention pooling as described under Section~\ref{sec:attn_pooling}.
%RescoreBERT and RescoreGPT refers to models under training scheme proposed in Section~\ref{sec:pll}.
%RescoreGPT-LL and RescoreBERT-PLL refers to the training scheme under Section~\ref{sec:ll}.

All configurations of discriminative training improves upon the log-likelihood baseline systems.
GPT-2 with MWER fine-tuning provides a relative WER improvement of up-to 5.9\%.
Similarly, BERT with PLL based MWER fine-tuning provides up-to 8.49\%.
It is interesting to note that the gap between GPT-2 and BERT after MWER training is increased from 1.26\% to 4.01\% relative.
This suggests that the discriminative training can exploit the bi-directional information more efficiently.
In addition to the MWER loss objective, we also experiment using linear combination of MWER loss and cross-entropy (CE) loss ($\text{MWER} + \alpha \times \text{CE}$) to stabilize training as suggested in \cite{gandhe2020audio,chiu2018state} ($\alpha=0.01$ tuned over $\{0.01,0.1,1\}$).
While we did not observe significant gains in combining the cross-entropy objective along with the MWER loss for GPT-2, we notice improvements in case of the BERT.

Further, RescoreGPT model gives up-to 1.4\% improvements over the baseline GPT-2 model.
RescoreBERT \cite{xu2022rescorebert} gives relative improvements of up-to 7.21\% in comparison with the baseline BERT model.
This finding suggests (i) using the last-token embedding for rescoring in-case of causal language models is less effective, (ii) bi-directional encoding and [CLS] representation of BERT plays an important role in discriminative training.
This also suggests that including sequence probability in MWER computation is crucial in case of generative, causal language models like GPT-2.
For attention pooling variants, in both the cases, i.e., RescoreBERT and RescoreGPT, we do not observe significant gains.

%Next, using sequence probability for computing the MWER loss, as described in Section~\ref{sec:ll}, gives upto 5.9\% improvement over the GPT-2 baseline (see RescoreGPT-LL in Table~\ref{tab:results}).
%In our experiments, we did not observe significant gains in combining the cross-entropy objective along with the MWER loss.

%Comparing RescoreBERT with RescoreGPT-LL, we find RescoreBERT outperforms by a relative 2.67\% on test-clean and 1\% on test-other.
%The difference can be attributed to bi-directional encoding.
%It is interesting to note that the MWER training (RescoreBERT) can exploit bi-directional encoding to a larger extent (2.67\% rel. improvement) compared to without MWER training (1.26\% rel. improvement). 
%It is interesting to note that the advantages with bi-directional encoding is minimal even after discriminative training).
%This makes MWER trained GPT-2 models a viable and attractive alternative to BERT for ASR rescoring.

We also perform latency profiling to demonstrate the advantages and disadvantages of each of the explored architectures.
Table \ref{tab:results} shows model rescoring latency for ten random hypothesis with fixed length of 64 tokens. 
Rescoring latency was computed using pytorch in eager execution mode with a single NVIDIA T4 GPU. 
As expected, using PLLs from BERT model to rescore the hypothesis takes two magnitudes more time than other rescoring models. 
This is essentially due to the need to evaluate the model for each token. 
GPT-2 has roughly 40\% higher latency than the pooled counterparts (RescoreGPT and RescoreBERT). 
This added latency can be attributed to the evaluation of the vocabulary softmax demonstrating that predicting the score directly can cut off the latency significantly even for 100 million parameter models. 
It is also worthwhile to note that some portion of the GPT model's latency could be reduced by inferring the model during the 1st pass decoding itself, which cannot be done with any bidirectional model (e.g. BERT). 

\section{Conclusion}
\label{sec:conclusion}
In this work, we proposed and explored several discriminative training techniques (based on MWER criterion) in application to ASR second-pass rescoring designed for pre-trained language models, particularly GPT-2 and BERT.
We explored three configurations (i) typical sequence probability based MWER loss, (ii) tuning last token embedding with MWER loss, and (iii) attention pooling of all the output embeddings.
Experiments were conducted using publicly available dataset (LibriSpeech) and models.
Comparisons are carried out to assess the role of bi-directional encoding and its relevance to discriminative training.
Our results suggests that using the last token embedding of GPT-2 model is not as effective as using [CLS] token in BERT models.
However, we demonstrate that with sequence probability based MWER training of GPT-2 model, the gap is much closer to the BERT counterpart.
We find that discriminative training helps exploit the bidirectional information in a better way for rescoring.
Clear advantages in terms of latency is demonstrated for pooling techniques with trade-offs for WER.
In future, we plan to scale the size of rescoring models and check the applicability of zero-shot prompting with discriminative models.

\bibliographystyle{IEEEbib}
\bibliography{strings,refs}

\end{document}